\documentclass{llncs}
\usepackage[utf8]{inputenc}
\usepackage{amssymb}
\usepackage{graphicx}
\usepackage{soul}
\usepackage{color}
\begin{document}
\title{Curious Minds Wonder Alike: Studying Multimodal Behavioral Dynamics to Design Social Scaffolding of Curiosity}
\author{Tanmay Sinha \and Zhen Bai \and Justine Cassell \\School of Computer Science, Carnegie Mellon University, USA \\ \{tanmays, zhenb, justine\} @ cs.cmu.edu}
\institute{}
\maketitle
\vspace{-0.5cm}
\begin{abstract}
Curiosity is the strong desire to learn or know more about something or someone. Since learning is often a social endeavor, social dynamics in collaborative learning may inevitably influence curiosity. There is a scarcity of research, however, focusing on how curiosity can be evoked in group learning contexts. Inspired by a recently proposed theoretical framework \cite{sinhaectel1} that articulates an integrated socio-cognitive infrastructure of curiosity, in this work, we use data-driven approaches to identify fine-grained social scaffolding of curiosity in child-child interaction, and propose how they can be used to elicit and maintain curiosity in technology-enhanced learning environments. For example, we discovered sequential patterns of multimodal behaviors across group members and we describe those that maximize an individual's utility, or likelihood, of demonstrating curiosity during open-ended problem-solving in group work. We also discovered, and describe here, behaviors that directly or in a mediated manner cause curiosity related conversational behaviors in the interaction, with twice as many interpersonal causal influences compared to intrapersonal ones. We explain how these findings form a solid foundation for developing curiosity-increasing learning technologies or even assisting a human coach to induce curiosity among learners. 
\end{abstract}
\vspace{-0.8cm}

\section{Introduction and Motivation}
\vspace{-0.4cm}
Curiosity is an important metacognitive skill that arises from a strong desire for learning \cite{berlyne1960conflict} and leads to knowledge acquisition through coming to one’s own understanding, rather than ``being told" or ``instructed". While there is an increasing emphasis on the educational benefits of learning in groups, as co-constructivism and collaborative learning theories argue that knowledge is jointly constructed through social interactions \cite{chi2014icap}, existing research on curiosity mainly focuses on investigating its cognitive mechanisms at an individual level, and often conceives curiosity as an inherently individual and stable disposition toward seeking novelty and approaching unfamiliar stimuli \cite{grossnickle2016disentangling}. Ignoring social factors in evoking curiosity may prevent us from designing effective forms of support in learning environments (technological or not), because in group work the behaviors of each member (both what they say and what they do) affect the curiosity of others \cite{forsyth2009group}. Prior learning sciences literature on the social and technological dimensions of scaffolding emphasizes that ``scaffolds are not found in software but are functions of processes that relate people to performances in activity systems over time" \cite{pea2004social}. It is therefore important to investigate the dynamics of these fine-grained processes as they happen spontaneously.

The theoretical motivation for studying these ``multimodal behavioral dynamics" (as we will call them) in order to better understand how to design for social scaffolding of curiosity stems from a fundamental psychological question - what causes variations in the curiosity level of children as they engage in open-ended collaborative problem-solving activities? Patterns of verbal and nonverbal behaviors comprise salient cues, and can provide valuable insights into how an individual's curiosity changes as they progress through the task. However, looking at summative measures (e.g - frequency of productive versus unproductive learning behaviors) alone will not suffice in understanding how curiosity arises and disappears over time. We believe that studying the social scaffolding of curiosity therefore requires examining sequential behavioral patterns that co-occur with – or just before - high curiosity moments, and then explicitly modeling the precise nature of causal relationships among these interpersonal patterns. Prior work on studying curiosity has not adequately addressed these behavioral dynamics. Even research that has looked at the effect of peers on curiosity has looked into mostly dyadic contexts rather than small group, used a limited strategy repertoire for eliciting curiosity-related behavior based on theory rather than empirical data, and subjectively assessed success of those strategies post-hoc using questionnaires \cite{graesser1995collaborative,gordon2015can,wu2013modeling}.

In this paper, then, we look at the social scaffolding of curiosity in detail, based on audio and video data of groups of elementary and middle school students engaged in informal learning. A subset have been coded for ground truth curiosity (see below for an explanation of what we mean) and a wide range of multimodal behaviors, using a mix of manual and semi-automated procedures. These behaviors are specified in the theoretical framework of curiosity, which we proposed and empirically validated in other work \cite{sinhaectel1} by articulating the underlying functions of these behaviors in contributing to curiosity in group learning. Building on this theoretical framework, we here address the research question of how to elicit these behaviors. To that end, we first look into sequential patterns of behaviors across group members that maximize an individual's curiosity within every one minute time frame. These sequential patterns inform \textit{what} behaviors to elicit in increasing or maintaining curiosity level of the target subject, based on the behavior trajectories recognized so far. We then study causal relationship between these behaviors to establish strategies of \textit{how} to elicit certain behaviors. The main contribution of this work is novel data-driven behavioral heuristics that we discover for enabling the design of supportive and responsive learning environments that can foster curiosity. In remainder of this paper, we first describe methods including data collection, annotation and analyses in section 2, followed by discussion of results in section 3. We end with implications for designing learning technologies and conclusion in sections 4 and 5. 
\vspace{-0.45cm}

\section{Method}
\vspace{-0.45cm}
In preparation for analyses of sequential behavioral patterns, we used the same annotated dataset annotated described in \cite{sinhaectel1}, which we summarize here as well. We then describe the detailed rationale behind our multimodal data analyses.
\vspace{-0.46cm}
\subsection{Data Collection}
\vspace{-0.19cm}
Audio and video data was collected for 12 groups of children (aged 10-12, 3-4 children per group, 44 in total) engaged in a hands-on activity commonly used in informal learning contexts - collaboratively build a Rube Goldberg machine (RGM). A RGM includes building chain reactions that are to be triggered automatically for trapping a ball in a cage. This paper describes fine-grained analyses of the first 30 minutes (out of 35-40 minutes given each group) of the RGM task for half of the sample; that is, 22 children across 6 groups.
\vspace{-0.5cm}

\subsection{Data Annotation}
\vspace{-0.25cm}
\subsubsection{Ground Truth Curiosity:}
Person perception research has demonstrated that judgments of others based on brief exposure to their behaviors is an accurate assessment of interpersonal dynamics \cite{ambady1992thin}. We used the Amazon MTurk to obtain ground truth for curiosity via such a thin-slice approach, using the definition ``curiosity is a strong desire to learn or know more about something or someone", and a rating scale comprising 0 (not curious), 1 (curious) and 2 (extremely curious). Amazon MTurk is a crowdsourcing platform that allows online workers to complete tasks that computers are currently unable to do, for a monetary payment. Our previous research has successfully deployed thin-slice coding for other social phenomena like rapport using this platform \cite{sinha2015we}. Four naive raters annotated every 10 second slice of videos of the interaction for each child presented to them in randomized order. We post-processed the ratings by removing those raters who used less than 1.5 standard deviation time compared to the mean time taken for all rating units (HITs). We then computed a single measure of Intraclass correlation coefficient (ICC) for each possible subset of raters for a particular HIT, and then picked ratings from the rater subset that had the best reliability for further processing. Finally, inverse-based bias correction \cite{kruger2014axiomatic} was used to account for label overuse and underuse, and to pick one single rating of curiosity for each 10 second thin-slice. The average ICC was 0.46.
\vspace{-0.15cm}

\subsubsection{Verbal and Non-verbal Behaviors:}
\vspace{-0.35cm}
We used semi-automatic (machine learning $+$ human judgment) and manual (human judgment) annotation procedures to code 11 verbal behaviors of interest in our corpus that came from our review of prior research in psychology and learning sciences, and our hypotheses about how these behaviors fulfill putative functions of curiosity. In other work, we have described details of the coding procedure, empirical validation of these hypotheses, and confirmation of positive predictive relationships between these behaviors, functions (that, because they cannot be directly observed, were our latent variables) and thin-slice curiosity \cite{sinhaectel1}. Here, in Table 1, we provide a summarized description of the verbal behaviors of uncertainty, argument, justification, suggestion, question asking (on-task, social), idea verbalization, sharing findings, hypothesis generation, attitude/sentiment towards task (positive, negative) and evaluation (positive, negative) that were coded at the clause level, and agreement that was coded at the turn level. A clause contains a subject (a noun or pronoun) and a predicate (conjugated verb – that says something about what the subject is or does). During a full turn, a speaker holds the floor and expresses one or more interpretable clauses (propositions). Inter rater reliability (Krippendorf's alpha) for each of these annotations was above 0.7. It is important to note that the above
annotation categories are not mutually exclusive, and can co-occur. In addition
to these verbal behaviors, we also used automated visual analysis to construct
five facial-landmark feature groups corresponding to emotional expressions that
provide evidence for the presence of affective states of joy, delight, surprise,
confusion and 
flow. More details are described in \cite{sinhaectel1}.

\vspace{-0.5cm}

\begin{table*} [hp]
\centering

\scalebox{0.72}{
\begin{tabular}{|p{3.7cm}|p{13cm}|} \hline
{\bf Verbal Behavior}
& {\bf Definition and Corpus Examples} \\ \hline
1. Uncertainty & Lack of certainty about ones choices or beliefs, and is verbally expressed by language that creates an impression that something important has been said, but what is communicated is vague, misleading, evasive or ambiguous. e.g - {\em``well maybe we should use rubberbands on the foam pieces", ``wait do we need this thing to funnel it through?"} \\ \hline
2. Argument & A coherent series of reasons, statements, or facts intended to support or establish a point of view. e.g - {\em ``no we got to first find out the chain reactions that it can do", ``wait, but anything that goes through is gonna be stuck at the bottom"} \\ \hline
3. Justification & The action of showing something to be right or reasonable by making it clear. e.g - {\em ``oh we need more weight to like push it down", ``wait with the momentum of going downhill it will go straight into the trap"} \\ \hline
4. Suggestion & An idea or plan put forward for consideration. e.g - {\em ``you could kick a ball to kick something", ``you are adding more weight there which would make it fall down"} \\ \hline
5. Question Asking & Asking any kind of questions related to the task (e.g - {\em ``so what's gonnna..what will happen like after the balls gets into the cup?", ``why do we need to make it that high?", ``do you want to build something like a chain reaction or something like that?"}) or non-task relevant (e.g - {\em ``do you two go to the same school?", ``who else watched the finale of gravity falls?"}) aspects of the social interaction \\ \hline 
6. Idea Verbalization & Explicitly saying out an idea, which can be just triggered by an individual’s own actions or something that builds off of other peer’s actions. e.g - {\em ``yeah that ball isn't heavy enough", ``so it's like tilted a bit up so it catches it instead of tilted down"} \\ \hline
7. Sharing Findings & An explicit verbalization of communicating results, findings and discoveries to group members during any stage of a scientific inquiry process. e.g - {\em ``look how I'm gonna see I'm gonna trap it", ``look I made my pillar perfect"} \\ \hline
8. Hypothesis Generation & Expressing one or more different possibilities or theories to explain a phenomenon by giving relation between two or more variables. e.g - {\em ``we could use scissors to cut off the baby's head which would cause enough friction", ``okay we need to make it straight so that the force of hitting it makes it big"} \\ \hline
9. Task Sentiment & A view of or attitude (emotional valence) toward a situation or event; an overall opinion towards a subject matter. We were interested in looking at positive (e.g - {\em ``oh it's the coolest cage I've ever seen, I'd want to be trapped in this cage", ``ok so I'm gonna try to find out a way for the end to make this one go and fall"}) or negative attitude (e.g - {\em ``I'm getting very mad at this cage",``but I don't know how to make it better"}) towards the task that students were working on \\ \hline 
10. Evaluation  &  Characterization of how a person assesses a previous speaker’s action and problem-solving approach. It can be positive (e.g - {\em ``oh that's a pretty good idea - that was a good idea",``let's make this thing elevated and make it go down"}) or negative (e.g - {\em ``oh wait this doesn't- you're not pushing anything over here", ``no it can't go like that otherwise it will be stuck"}) \\ \hline
11. Agreement & Harmony or accordance in opinion or feeling; a position or result of agreeing. e.g - {\em ``But we need to have like power, and weight too" {\em (Quote)} --- ``Yeah we need more weight on this side" {\em (Response)}, ``And we put the ball in here..I hope it still works, and it goes..so it starts like that, and then we hit it" {\em (Quote)} --- ``Ok that works" {\em (Response)}} \\ \hline 
\end{tabular}}
\caption{Definition \& Examples of Curiosity-related verbal behavior coded. Detailed coding scheme can be found at http://tinyurl.com/codingschemecuriosity}\label{tab:1}
\end{table*}
\vspace{-1.6cm}

\subsection{Multimodal Data Analyses}
\vspace{-0.25cm}
We now describe our data-driven approach for discovering behavioral sequences that maximize curiosity and causal relationships between these behaviors.
\vspace{-0.28cm}

\subsubsection{Temporal Behavioral Relationships that Maximize Curiosity:}
\vspace{-0.2cm}
To discover the temporal relationships among multimodal behaviors that maximize curiosity, we needed to specify how these behavioral states change over time. We therefore used sequential pattern mining approaches to find productive high-curiosity conversational episodes in the group interaction. Traditionally, the selection of such interesting sequences is based on the frequency/support framework, where sequences of high frequency are treated as significant. However, this often leads to  many patterns being identified, most of which are may not be informative enough for choosing precise forms of scaffolding. Some sequential patterns, despite occurring rarely (having frequencies lower than the given minimum support), might still be useful since they co-occur with episodes of high individual curiosity. On the contrary, there might be other sequential behavioral patterns that occur frequently, but mostly co-occur with episodes of low individual curiosity. This motivated our current approach of incorporating utility in the classical sequential pattern mining framework. Our objective was to find what sequence of group member's behaviors maximize an individual's curiosity. 

Towards this end, we leveraged the USpan algorithm \cite{yin2012uspan}, which uses lexicographic quantitative sequence tree to extract the complete set of high utility sequences, and includes efficient concatenation mechanisms and pruning strategies for calculating the utility of a node and its children. Formally, in our work, we represented an input behavioral sequence using 6 itemsets $X_1, X_2, ... X_6$, where each itemset represented an unordered set of distinct co-occurring behaviors from group members within a 10 second span, and therefore each input sequence spanned one minute. Every behavior displayed by a group member in each itemset was associated with an additional utility value, which we defined as the ground truth thin-slice curiosity for that particular group member for the corresponding 10 second slice. For each group, we ran multiple passes of the USpan algorithm, varying the objective function each time to be the overall curiosity of each individual group member within the minute span. The overall utility $O$ of curiosity of a sequential behavioral pattern $S$ was the sum of utilities associated with $S$ in each of the input sequences where it appeared. The final output of USpan algorithm in each pass therefore comprised all high utility sequential patterns above an overall threshold utility value of $O$. 
\vspace{-0.26cm}

\subsubsection{Social Influence of Curiosity-related Behaviors:}
\vspace{-0.26cm}
To examine how social interaction evoked curiosity, we needed to find the interdependence among behavioral signals at a fine-grained level. In many situations of interest, symmetric measures of behavioral coordination aren’t satisfactory to tear apart which signal is coordinating towards which. In our work, we therefore leveraged the notion of causal influence proposed by Granger \cite{ding200617}, which states that if the prediction of one time series could be improved by incorporating the knowledge of a second one, or, if variance of the autoregressive prediction error of the first time series at the present time is reduced by inclusion of past measurements from the second time series, then the second series is said to have a causal influence on the first. For three or more simultaneous time series, a pairwise analysis can be performed to reduce the problem to a bivariate problem, the limitation however being that the causal relation between any two of the series may be direct, mediated by a third one, be a combination of both. This situation can be addressed by the technique of conditional Granger causality. 

Formally, to determine whether causal influence of behavioral time series Y on X was mediated by Z, we created two ordinary least square auto-regressive models - (i) Restricted (RR), where we predicted X using past values of X and Z, (ii) UnRestricted (UR), where we predicted X using past values of X, Y and Z. The conditional granger causality magnitude (G-ratio) of Y influencing X, given Z (Y$\rightarrow$X$|$Z) =  log (variance(Residual$_{RR}$)/variance(Residual$_{UR}$)), which is essentially ratio of the log of variance of errors in the restricted and unrestricted regression. If G-ratio $<=$ 0, no further improvement of X can be expected by including past measurements of Y (full mediation). If G-ratio is $>$ 0, there is still a direct causal influence component from Y to X, and the inclusion of past measurements of Y in addition to that of X and Z results in better predictions of X (partial mediation). Maximum lag length was set to 6 (we looked back at most 6*10=60 seconds in the behavioral time series X, Y and Z), and the optimal lag length $M$ was the one that minimized the Bayesian Information Criterion (BIC) obtained by fitting the restricted and unrestricted regression models to the data. Statistical significance was computed using an F-test under the null hypothesis that one time series does not granger cause the other, where $F(M,n-k-1)$ = ((Sum of Square Residual$_{RR}$ - Sum of Square Residual$_{UR}$) * (n-k-1))/((Sum of Square Residual$_{UR}$)* $M$), where $n$ is the number of observations, $k$ is the number of explanatory variables in the unrestricted regression, and $n-k-1$ refers to the residual degrees of freedom. We acknowledge that our notion of influence is based on cause-effect relations with constant conjunctions and is only a limited view of causation, and we invite future work to build upon this approach.
\vspace{-0.5cm}

\section{Results and Discussion}
\vspace{-0.4cm}
This section discusses representative behavioral patterns and causal relationships that resulted from our described analyses in section 2.3. To reiterate, our goal behind running these analyses was to inform the social scaffolding of curiosity by discovering \textit{what} behaviors to elicit in increasing or maintaining curiosity level of the target subject based on the behavior trajectories recognized so far, and then discovering strategies of \textit{how} to elicit these particular behaviors.
\vspace{-0.48cm}

\subsection{Temporal Behavioral Relationships that Maximize Curiosity}
\vspace{-0.25cm}
We synthesized representative sequential behavioral patterns across group members with high utility of individual curiosity by selecting those patterns that had a curiosity utility higher than 35 (where 35 was the average utility across all patterns discovered). For clarity, we explain these patterns along 5 themes based on the behaviors involved (Table.~\ref{tab:1}). Each pattern spans a total of 60 seconds, and comprises multiple co-occurring behavioral itemsets. Each of these individual itemsets, although unordered, is linked sequentially across time with a subsequently occurring itemset. For e.g, a pattern $B_a(other), B_b(other) {\Large \twoheadrightarrow} B_c(own)$ means that a behavioral itemset comprising behaviors A and B done by a different group member within a 10 second span are followed by a behavioral itemset comprising behavior C done by the target individual within the one minute span, and the pattern maximizes curiosity of this target individual.

Group 1 comprises patterns following the general theme of {\bf ideation} that are linked to high curiosity. In this group, {\em justification} comes up as a frequently co-occurring and contingent behavior with {\em idea verbalization} and together maximizes the utility of curiosity. {\em Justification} attempts to establish an idea's validity by linking it to evidence. This in turn helps identify errors in group problem solving, and clarifies relationships among task subcomponents to trigger creation of new ideas \cite{chen2012social}. For example, in the RGM task, group members often initially start working on different parts needed to assemble a complete RGM, and subsequently engage in justifying why and how their solution sub-pieces can be integrated. We also see that contingent occurrences of {\em idea verbalization} done by group members maximizes curiosity. Prior work \cite{paulus2003enhancing} has posited that group members may build on one another's diverse perspectives to create new ideas via underlying mechanisms such as activation of related concepts (sparked ideas), engagement into putting together pieces of a solution (jigsaws) and creative misinterpretations of incorrect ideas.

\vspace{-0.5cm}

\begin{table*} [hp]
\centering
\scalebox{0.85}{
\begin{tabular}{|p{12.5cm}|} \hline
{\bf Corpus Examples of Sequential Behavioral Patterns [Utility of Curiosity]} \\ \hline

{\scriptsize {\bf Theme 1: involving Justification (J), Idea verbalization (IV)}} \newline
1. IV(own) {\Large $\twoheadrightarrow$} J(own), IV(own) {\Large $\twoheadrightarrow$} J(own), IV(own) {\Large $\twoheadrightarrow$} J(own) \textbf{[129]}\\ 
2. J(own) {\Large $\twoheadrightarrow$} J(own), IV(own) {\Large $\twoheadrightarrow$} IV(own) {\Large $\twoheadrightarrow$} Confusion (other) \textbf{[120]}\\ 
3. J(other) {\Large $\twoheadrightarrow$} J(own), IV(own) {\Large $\twoheadrightarrow$} J(own) \textbf{[108]}\\ 
4. J(own), J(other) {\Large $\twoheadrightarrow$} J(own) {\Large $\twoheadrightarrow$} J(own) \textbf{[94]}\\ 
5. J(own), IV(own) {\Large $\twoheadrightarrow$} J(own) {\Large $\twoheadrightarrow$} J(other) \textbf{[92]}\\
6. J(other) {\Large $\twoheadrightarrow$} J(other) {\Large $\twoheadrightarrow$} J(own) \textbf{[67]}\\
\hline 

{\bf {\scriptsize Theme 2: involving Neg/Pos Evaluation (NE/PE), Justification (J), Idea verbalization (IV)}}  \\ 
1.  PE (own), J(own) {\Large $\twoheadrightarrow$} J(own)  \textbf{[80]} \\
2.  Confusion(other) {\Large $\twoheadrightarrow$} NE(other)  \textbf{[59]}\\ 
3. NE(other) {\Large $\twoheadrightarrow$} PE(own), J(own), IV(own) {\Large $\twoheadrightarrow$} J(own), Confusion (own)  {\Large $\twoheadrightarrow$}  PE(own), J(own) {\Large $\twoheadrightarrow$} J(own) \textbf{[55]}\\

\hline 

{\bf {\scriptsize Theme 3: involving Question asking Task (QAT), Justification (J), Idea verbalization (IV)}}  \\ 
1. Confusion(other), QAT(own) {\Large $\twoheadrightarrow$}  Confusion(other) {\Large $\twoheadrightarrow$} Confusion(other) \textbf{[53]}\\
2. J(other), IV(other) {\Large $\twoheadrightarrow$}  QAT(other) \textbf{[52]}\\
3. Confusion(own) {\Large $\twoheadrightarrow$}  QAT(other) \textbf{[45]}\\
\hline 

{\bf {\scriptsize Theme 4: involving Suggestion (S), Idea verbalization (IV)}}  \\ 
1. Confusion(other), S(own), IV(own), Confusion(own) {\Large $\twoheadrightarrow$}  Confusion(other), IV(own), Confusion(own) {\Large $\twoheadrightarrow$} Confusion(other) {\Large $\twoheadrightarrow$} IV(own) \textbf{[67]}\\

\hline

{\bf {\scriptsize Theme 5: involving Positive Emotional states, Positive Task Sentiment (PTS)}}  \\ 
1. Joy(own) {\Large $\twoheadrightarrow$} Joy(own) \textbf{[80]}\\
2. Joy(own), Delight(other) {\Large $\twoheadrightarrow$} Joy(own) \textbf{[55]}\\
3.  Confusion (other)  {\Large $\twoheadrightarrow$} PTS(other) \textbf{[44]}\\
4. Joy(other) {\Large $\twoheadrightarrow$} Flow(own) \textbf{[42]}\\

\hline

\end{tabular}}
\caption{Salient sequential behavioral pattern groups that maximize the utility of individual (own) curiosity for the pattern. Each pattern spans 60 seconds. Flow of time between subsequent behavioral itemsets within the pattern is depicted by {\Large $\twoheadrightarrow$}}\label{tab:1}
\end{table*}
\vspace{-1cm}

Group 2 comprises patterns following the general theme of {\bf evaluation} that are linked to high curiosity. {\em Positive evaluations} support correct information by showing solidarity, a desire for cooperation and expressing positive emotions. On the other hand, {\em negative evaluation} is often an expression of disagreement, where flaws are identified in a peer's problem-solving approach by being critical of or even dismissing the peer's idea. It results in conflict, and group members are motivated to reduce that conflict via discussion (increased involvement or commitment), by getting others to change (attempting an influence), seeking additional social support for the opinion held (adding new ideas that are consonant with one's own opinions) or by changing their own opinion. All these tactics for reducing opposing beliefs will involve sequential behaviors of {\em justification}, {\em idea verbalization} and further {\em evaluation} \cite{chiu2008flowing}, as we see in table 2. In addition, even if inaccurate, {\em negative evaluation} often stimulates the attention of group members, and therefore might help them consider more aspects of the task from different perspectives to aid in creation of new ideas indirectly \cite{orlitzky2001err}. The group dynamics literature provides complementary insights to explain the relationships between {\em evaluation} and the subsequent discussion trajectory - it suggests that {\em negative evaluations} made by some group members might be comparatively more tolerable than if they are made by others. Such {\em evaluations} are likely to be taken seriously (rather than being dismissed or overruled), and there will be a high motivation to consider and resolve the obstacle by engaging in reasoning together, which can trigger curiosity. This can happen, for instance, because of positive impressions of a group member held by others that accumulate as members contribute to progress of the group towards desired goals, or, if certain group members possess valuable personal characteristics \cite{cartwright1953group}.
\vspace{-0.05cm}

Group 3 comprises patterns following the general theme of {\bf closing knowledge gaps} \cite{loewenstein1994psychology} and that are linked to high curiosity. These comprise {\em question asking} behaviors that co-occur or are contingent with {\em confusion-related facial expressions}. Prior literature in socio-emotional learning \cite{d2012dynamics} has found {\em confusion} to be a key signature of cognitive disequilibrium, or, a state of uncertainty, and occurs when an individual faces contradictions or comes across novel stimuli, both of which are precursors of curiosity \cite{berlyne1960conflict}. In our work, we coded for questions belonging to specific task aspects such as how and why things work, what-if something affects or will affect something else, underlying mechanisms or causal factors of a process or observation in detail, and other general knowledge (e.g. fact, terms, classification, or other general information) as on-task questions \cite{luce2015science}. Such {\em on-task question asking} in group work, which reflects lacunae in understanding, reveals uncertainties in front of group members, and can be part of a think-aloud about the subject matter/specific scientific phenomenon/task that students are working on themselves. Think aloud in scientific inquiry helps monitor one's own thinking and understanding, and initiates meta-cognitive reflection to trigger awareness of knowledge gaps for engaging in further exploration. When tackling complex tasks in open-ended collaborative learning environments, thinking aloud together has been empirically shown to regulate co-construction of knowledge and lead to improvement in the ability to articulate collaborative reasoning processes \cite{hogan1999thinking,mockel2013thinking}. {\em On-task question} asking can also be part of a question asked to another group member regarding what they are working on, how they act and think, their opinions or requesting suggestions relating to the task. We find in our RGM corpus that when group members recognize problematic ideas or flaws in the chain-reaction sub-components made by a peer, they often ask questions to express these knowledge gaps and elicit more information. These questions invite further {\em idea verbalization}. 

Group 4 comprises patterns involving {\bf making suggestion} to other group members, where an idea, possible plan or action for others to consider is mentioned, or, an opinion about what other people should do and how they should act in a particular situation is offered. Making {\em suggestions} is an evidence that a shared conception of the problem has very likely been developed, and therefore the {\em suggestion} is geared towards engaging in cooperative effort to overcome the obstacle, and joint creation of new interpretations. Thus, at a fundamental level, it not only signals interest in other's work, but also a child’s anticipation to know whether the proposed idea will work or not (impact of the suggestion) and therefore find out the uncertain/unknown result. Engaging in these socio-cognitive processes of knowledge acquisition will spur an individual's curiosity, as is evident from the high utility sequential pattern shown in table 2.

Group 5 comprises the dynamics of {\bf positive emotional states} \cite{d2012dynamics} that maximize the utility of curiosity. {\em Delight} and {\em joy} denote the pleasure associated with discovering new ideas by oneself or other group members. Emotional expressions of {\em flow} point to spending time and effort in acquiring a solution. It is indicative of persistence in engaging in knowledge acquisition processes.
\vspace{-0.4cm}

\subsection{Social Influence of Curiosity-related Behaviors}
\vspace{-0.25cm}
To investigate social influence, we first ran the conditional granger causality algorithm separately for each group. We then synthesized similar causal behavioral influences across groups that were significant at 0.001 level of significance and averaged their G-ratios for presentation (Table 3 and 4). Overall, we found $\sim$2x higher number of significant interpersonal causal influence involving 2 or more group members (325) compared to intrapersonal causal influence (154). This strongly points towards why social scaffolding in group work is necessary, which corroborates with other work \cite{sinhaectel1}, as well as the precise way to provide it. We describe these significant causal influences at the interpersonal level along 4 themes and explain our interpretation of these results below (see table 3 and 4). 

Group 1 reflects the theme of {\bf behavioral contagion}, or the propensity for certain behavior exhibited by a group member to be repeated in close temporal proximity by others. The putative mechanism underlying this social phenomena might be entrainment, which in previous work we found had an impact on rapport and learning \cite{sinha2015we,sinha2015exploring}, or alternately, can also involve careful evaluation of conditions under which group members would be willing to be influenced. These conditions can involve looking at the motivational consequences of accepting or rejecting the influencing peer's behavior, such as the desire to receive reward or avoid punishment, desire to be like an admired person in the group (normative social influence), desire to abide by one’s values (establishing self-identity), desire to be correct (informational social influence), other group oriented desires (such as welfare of the group), or intrinsically rewarding consequences \cite{cartwright1953group}.

In particular, in table 3, we can see a significant causal influence of {\em uncertainty} expressed by one child on {\em uncertainty} of another child. Looking through the lens of group dynamics \cite{dornyei2003group}, closely contingent expressions of {\em uncertainty} from group members about similar (or related) aspects of the task is a signal of ``joint hardship", or the experience of common blocking points for the group to proceed in its task. This causal relationship has been posited to positively influence the social interaction, since individuals expressing uncertainty will subsequently engage in cooperative effort to overcome the cause of uncertainty, often enhancing acceptance and group attraction because of having coped with the hardship situation. Moreover, the hope of resolving uncertainty under joint effort will make children more eager to explore, in turn increasing their curiosity. In addition, we also see significant interpersonal causal influences along behavioral constructs such as {\em sharing findings}, {\em argument} and {\em social question asking} (see table 3). Such social questions reflect a general interest in gaining new social information about non-task relevant personal information and feelings, likes, dislikes, preferences from other group members \cite{litman2007dimensionality}. They are a motivator for joint exploratory behaviors since they increase group member familiarity, build interpersonal closeness and promote an unconditional positive regard towards group members \cite{dornyei2003group,rogers1983freedom}.

\vspace{-0.53cm}
\begin{table*} [hp]
\centering
\scalebox{0.84}{
\begin{tabular}{|p{13cm}|p{1.3cm}|} \hline
{\bf Social Influence (Direct)} & {\bf G-ratio} \\ \hline
{\bf {\scriptsize Theme 1: Contagion}} & \\ 1. Uncertainty (other) {\Large $\rightsquigarrow$} Uncertainty (own) & 0.687 \\
2. Sharing Findings (other) {\Large $\rightsquigarrow$} Sharing Findings (own) & 0.223 \\
3. Question Asking Social (other) {\Large $\rightsquigarrow$} Question Asking Social (own) & 0.379 \\  4. Argument (other) {\Large $\rightsquigarrow$} Argument (own) & 0.177 \\ \hline
{\bf {\scriptsize Theme 2: Constructive Controversy}} & \\ 1. Suggestion (other) {\Large $\rightsquigarrow$}  Argument (own) & 0.176\\
2. Argument (other) {\Large $\rightsquigarrow$} Idea Verbalization (own) & 0.160 \\ 
3. Argument (other) {\Large $\rightsquigarrow$} Negative Evaluation (own) & 0.138\\
4. Argument (other) {\Large $\rightsquigarrow$} Justification (own) & 0.131 \\ \hline
{\bf {\scriptsize Theme 3: Idea/View Refinement}} & \\ 1. Hypothesis generation (other) {\Large $\rightsquigarrow$}  Suggestion (own) & 0.256 \\ 
2. Question Asking Task (other)  {\Large $\rightsquigarrow$} Hypothesis generation (own) & 0.248 \\ 
3. Suggestion (other) {\Large $\rightsquigarrow$} Negative Evaluation (own) & 0.109 \\ 
4. Sharing Findings (other) {\Large $\rightsquigarrow$} Negative Evaluation (own) & 0.086 \\  \hline
{\bf {\scriptsize Theme 4: Supportive Responses}} & \\ 1. Uncertainty (other) {\Large $\rightsquigarrow$} Agreement (own) & 0.171 \\
2. Uncertainty (other) {\Large $\rightsquigarrow$} Suggestion (own) & 0.111 \\
3. Idea Verbalization (other) {\Large $\rightsquigarrow$} Positive evaluation (own) & 0.098 \\
4. Uncertainty (other) {\Large $\rightsquigarrow$} Hypothesis generation (own) & 0.086\\  \hline

\end{tabular}}
\caption{Salient examples of direct social influence ({\Large $\rightsquigarrow$}) along with corresponding conditional granger causality magnitudes (significant at 0.001 LOS)}\label{tab:2}
\end{table*}
\vspace{-1cm}

\vspace{-0.59cm}

\begin{table*} [hp]
\centering
\scalebox{0.84}{
\begin{tabular}{|p{13cm}|p{1.3cm}|} \hline
{\bf Social Influence (Fully Mediated)} & {\bf G-ratio} \\ \hline
{\bf {\scriptsize Theme 1: Constructive Controversy}} & \\ 
Argument (p1) {\Large $\rightsquigarrow$} Surprise (p2) {\Large $\rightsquigarrow$} Justification (p3)
 & 0.251 \\ \hline
 
{\bf {\scriptsize Theme 2: Idea/View Refinement}} & \\ 1. Hypothesis Generation (p1) {\Large $\rightsquigarrow$} Sharing Findings (p2) {\Large $\rightsquigarrow$} Suggestion (p3)
 & 0.399 \\ 2. Hypothesis Generation (p1) {\Large $\rightsquigarrow$} Sharing Findings (p2) {\Large $\rightsquigarrow$} Negative Evaluation (p3) & 0.250 \\
3.  Sharing Findings (p1)  {\Large $\rightsquigarrow$} Hypothesis Generation (p2) {\Large $\rightsquigarrow$} Idea Verbalization (p2)
 & 0.233 \\
4. Sharing Findings (p1)  {\Large $\rightsquigarrow$} Hypothesis Generation (p2) {\Large $\rightsquigarrow$} Justification (p2)
 & 0.167 \\ \hline
 
 {\bf {\scriptsize Theme 3: Supportive Responses}} & \\ Sharing Findings (p1)  {\Large $\rightsquigarrow$} Hypothesis Generation (p2) {\Large $\rightsquigarrow$} Positive Evaluation (p2)
 & 0.148 \\ \hline

\end{tabular}}
\caption{Salient examples of fully mediated social influence ({\Large $\rightsquigarrow$}) along with corresponding conditional granger causality magnitudes (significant at 0.001 LOS)}\label{tab:3}
\end{table*}
\vspace{-0.94cm}

Group 2 reflects the theme of {\bf constructive controversy} \cite{johnson2009energizing}, or group members' involvement in seeking out to reach an agreement when their ideas, conclusions and theories are incompatible with those of one another. Such constructive controversy, as instantiated in interpersonal behaviors such as {\em argument}, {\em negative evaluation} etc leads to an active search for additional perspectives to support correctness of one's own view. This is likely to improve the quality of group decision making by providing a medium through which problems can be aired and tensions released. This environment of self-evaluation and change will in turn encourage interest and curiosity among group members \cite{paletz2011intragroup}. For our corpus, some salient direct interpersonal causal influences from this group include those of {\em suggestion} on {\em argument}, {\em argument} on {\em idea verbalization}, {\em argument} on {\em negative evaluation} and {\em argument} on {\em justification} (see table 3). Additional fully mediated causal influences among behaviors in this group are shown in table 4, where we find {\em sharing findings} fully mediates the causal influence of {\em hypothesis generation} on {\em suggestion/negative evaluation}. In addition, {\em hypothesis generation} fully mediates the causal influence of {\em sharing findings} on {\em idea verbalization/justification}.

Group 3 reflects the theme of {\bf refining a group member's ideas or views}. This can be seen via direct interpersonal causal influences of {\em hypothesis generation} on {\em suggestion}, {\em task question asking} on {\em hypothesis generation}, {\em suggestion} on {\em negative evaluation} and {\em sharing findings} on {\em negative evaluation} in table 3. Prior work has posited that such {\em negative evaluation}, as a common expression of disagreement referring to epistemic (task) content, will enhance an individual's curiosity because of enhancement of perceived contribution of the peer \cite{darnon2007dealing}. Additional fully mediated causal influences among behaviors in this group are shown in table 4, where we find that the causal influence of {\em argument} made by person A on {\em justification} done by person B is fully mediated by an emotional expression of {\em surprise} from a third group member person C.

Group 4 reflects the theme of {\bf supportive responses to uncertainty}, which are more likely when one's peers either share the uncertainty or at least consider it warranted, reasonable, or legitimate \cite{jordan2014managing}. In particular, for our corpus, some salient direct interpersonal causal influences include those of {\em uncertainty} on {\em agreement/suggestion/hypothesis generation}, and {\em idea verbalization} on {\em positive evaluation} (see table 3). Additional fully mediated causal influence among behaviors in this group are shown in table 4, where we find that the causal influence of {\em sharing findings} by person A on {\em positive evaluation} made by person B is fully mediated by {\em hypothesis} generated by person B.
\vspace{-0.43cm}

\section{Implications for Designing Learning Technologies}
\vspace{-0.4cm}
In spite of its critical link with learning, curiosity is often found to decrease with age and schooling, partially because of prevalence of test-oriented education strategies that follow from educational policies such as the ``common core"\cite{porter2011common}. This effect is even more pronounced in inner city classrooms with limited teaching resources that are constantly under great pressure to adhere to academic standards. Understanding how to design computer support to raise and sustain curiosity will make this important metacognitive skill more accessible to students from diverse socioeconomic backgrounds. In this paper then, we claim that such forms of computer support should be equipped with fine-grained understanding of the unfolding behavioral trajectory, to allow for detection of behaviors belonging to a larger sequential pattern that maximizes the utility of curiosity for a target learner. Our work in the first part of this paper can aid in development of data-driven heuristics for providing a principled way of choosing the kind of support to be provided (given the observed behavior trajectories). However, since not all productive conversational behaviors that maximize the utility of curiosity in human-human interaction might occur naturally in interactions between human and a learning technology, it might be worthwhile to make some arrangements for the appearance of such behaviors. We can then leverage insights gained from second part of the work presented in this paper to decide an action (behavior) to be performed by a learning technology that will cause/trigger a particular behavioral change in a peer. 

Investigation of social influence of curiosity-related behaviors provides a simple, yet elegant solution to an important and fundamental research question in human perception and reasoning - given a desired mental state change (curiosity), how can a learning technology (for example, in the form of a pedagogical agent) act to cause that mental state change in a human. For example - let's suppose we have the sequential behavioral pattern of: Task Question Asking(person 2) {\Large $\twoheadrightarrow$} Uncertainty(person1) that maximizes the utility of curiosity of person 1. On perceiving that person 2 has asked a task-related question, and person 1 is passive in subsequent time steps, the social influence knowledge database can be consulted and the specific causal influence rule of: Uncertainty (other) {\Large $\rightsquigarrow$} Uncertainty (own) can be picked by a pedagogical agent to verbalize an expression of uncertainty about some aspect of the task that was related to the question asked by person 2, along with (maybe) asking person 1's opinion about the same. This is likely to capture person 1's attention, who might express uncertainty about similar aspects of the task. Such shared uncertainty might make person 1 eager to reduce their knowledge gap by engaging in joint exploration, in turn maximizing their curiosity. Furthermore, since data-driven approaches cannot capture the exhaustive set of productive social interaction practices that educators have been using for raising children's curiosity in different learning settings (e.g - promoting risk taking by rewarding exploration of diverse solutions, helping group members find causal relationships between processes by asking them to make an explicit link between learning representations) \cite{spektor2013science,correnti2015improving}, we must acknowledge that results derived from this research can be augmented with those top-down strategies to provide complementary benefits to a learner.  
\vspace{-0.45cm}

\section{Conclusion}
\vspace{-0.35cm}
In this work, we looked at sequential patterns of multimodal behaviors across group members that maximize an individual's utility of curiosity when learning in social contexts. To provide rich forms of social scaffolding for fostering curiosity, we further investigated direct and mediated interpersonal causal influences that can be used to trigger particular productive conversational behaviors in the interaction. These results draw on various theoretical lenses in learning sciences and the social psychology of group dynamics, as well as results from our analyses of small group informal learning. We believe that such a fine-grained theoretical understanding of the construct of curiosity holds the key to combating its absence in collaborative learning settings by leveraging simple, yet powerful insights that we gain from analytical approaches outlined in this work. The underlying rationale is applicable more generally for developing computer support for other metacognitive skills as well. Our larger vision is to develop socially-aware learning technologies \cite{zhao2016socially} that can bring back an individual's curiosity, maintain the momentum ignited by it, and help individuals engage in task-completion by pooling interpersonal resources when working in a group, motivated by their intrinsic interest. Through the design of such learning technologies and confirming their effectiveness, we also hope to provide additional pedagogical instructions for school teachers to help children with diverse socio-economical background develop knowledge-seeking skills driven by intrinsic curiosity.
\vspace{-0.4cm}

\bibliographystyle{splncs03}
\bibliography{splncs} 

\begin{thebibliography}{10}
\providecommand{\url}[1]{\texttt{#1}}
\providecommand{\urlprefix}{URL }

\bibitem{ambady1992thin}
Ambady, N., Rosenthal, R.: Thin slices of expressive behavior as predictors of
  interpersonal consequences: A meta-analysis. (1992)

\bibitem{berlyne1960conflict}
Berlyne, D.E.: Conflict, arousal, and curiosity.  (1960)

\bibitem{cartwright1953group}
Cartwright, D.E., Zander, A.E.: Group dynamics research and theory.  (1953)

\bibitem{chen2012social}
Chen, G., Chiu, M.M., Wang, Z.: Social metacognition and the creation of
  correct, new ideas: A statistical discourse analysis of online mathematics
  discussions. Computers in Human Behavior  28(3),  868--880 (2012)

\bibitem{chi2014icap}
Chi, M.T., Wylie, R.: The icap framework: Linking cognitive engagement to
  active learning outcomes. Educational Psychologist  49(4),  219--243 (2014)

\bibitem{chiu2008flowing}
Chiu, M.M.: Flowing toward correct contributions during group problem solving:
  A statistical discourse analysis. The Journal of the Learning Sciences
  17(3),  415--463 (2008)

\bibitem{correnti2015improving}
Correnti, R., Stein, M.K., Smith, M.S., Scherrer, J., McKeown, M., Greeno, J.,
  Ashley, K.: Improving teaching at scale: Design for the scientific
  measurement and learning of discourse practice. Socializing Intelligence
  Through Academic Talk and Dialogue. AERA  (2015)

\bibitem{darnon2007dealing}
Darnon, C., Doll, S., Butera, F.: Dealing with a disagreeing partner:
  Relational and epistemic conflict elaboration. European Journal of Psychology
  of Education  22(3),  227--242 (2007)

\bibitem{ding200617}
Ding, M., Chen, Y., Bressler, S.L.: 17 granger causality: basic theory and
  application to neuroscience. Handbook of time series analysis: recent
  theoretical developments and applications  437 (2006)

\bibitem{dornyei2003group}
D{\"o}rnyei, Z., Murphey, T.: Group dynamics in the language classroom. Ernst
  Klett Sprachen (2003)

\bibitem{d2012dynamics}
D’Mello, S., Graesser, A.: Dynamics of affective states during complex
  learning. Learning and Instruction  22(2),  145--157 (2012)

\bibitem{forsyth2009group}
Forsyth, D.R.: Group dynamics. Cengage Learning (2009)

\bibitem{gordon2015can}
Gordon, G., Breazeal, C., Engel, S.: Can children catch curiosity from a social
  robot? In: Proceedings of the Tenth Annual ACM/IEEE International Conference
  on Human-Robot Interaction. pp. 91--98. ACM (2015)

\bibitem{graesser1995collaborative}
Graesser, A.C., Person, N.K., Magliano, J.P.: Collaborative dialogue patterns
  in naturalistic one-to-one tutoring. Applied cognitive psychology  9(6),
  495--522 (1995)

\bibitem{grossnickle2016disentangling}
Grossnickle, E.M.: Disentangling curiosity: Dimensionality, definitions, and
  distinctions from interest in educational contexts. Educational Psychology
  Review  28(1),  23--60 (2016)

\bibitem{hogan1999thinking}
Hogan, K.: Thinking aloud together: A test of an intervention to foster
  students' collaborative scientific reasoning. Journal of Research in Science
  Teaching  36(10),  1085--1109 (1999)

\bibitem{johnson2009energizing}
Johnson, D.W., Johnson, R.T.: Energizing learning: The instructional power of
  conflict. Educational Researcher  38(1),  37--51 (2009)

\bibitem{jordan2014managing}
Jordan, M.E., McDaniel~Jr, R.R.: Managing uncertainty during collaborative
  problem solving in elementary school teams: The role of peer influence in
  robotics engineering activity. Journal of the Learning Sciences  23(4),
  490--536 (2014)

\bibitem{kruger2014axiomatic}
Kruger, J., Endriss, U., Fern{\'a}ndez, R., Qing, C.: Axiomatic analysis of
  aggregation methods for collective annotation. In: Proceedings of the 2014
  international conference on Autonomous agents and multi-agent systems. pp.
  1185--1192. International Foundation for Autonomous Agents and Multiagent
  Systems (2014)

\bibitem{litman2007dimensionality}
Litman, J.A., Pezzo, M.V.: Dimensionality of interpersonal curiosity.
  Personality and Individual Differences  43(6),  1448--1459 (2007)

\bibitem{loewenstein1994psychology}
Loewenstein, G.: The psychology of curiosity: A review and reinterpretation.
  Psychological bulletin  116(1), ~75 (1994)

\bibitem{luce2015science}
Luce, M.R., Hsi, S.: Science-relevant curiosity expression and interest in
  science: an exploratory study. Science Education  99(1),  70--97 (2015)

\bibitem{mockel2013thinking}
Mockel, L.J.: Thinking aloud in the science classroom: Can a literacy strategy
  increase student learning in science?  (2013)

\bibitem{orlitzky2001err}
Orlitzky, M., Hirokawa, R.Y.: To err is human, to correct for it divine a
  meta-analysis of research testing the functional theory of group
  decision-making effectiveness. Small Group Research  32(3),  313--341 (2001)

\bibitem{paletz2011intragroup}
Paletz, S.B., Schunn, C.D., Kim, K.H.: Intragroup conflict under the
  microscope: Micro-conflicts in naturalistic team discussions. Negotiation and
  Conflict Management Research  4(4),  314--351 (2011)

\bibitem{paulus2003enhancing}
Paulus, P.B., Brown, V.R.: Enhancing ideational creativity in groups. Group
  creativity: Innovation through collaboration pp. 110--136 (2003)

\bibitem{pea2004social}
Pea, R.D.: The social and technological dimensions of scaffolding and related
  theoretical concepts for learning, education, and human activity. The journal
  of the learning sciences  13(3),  423--451 (2004)

\bibitem{porter2011common}
Porter, A., McMaken, J., Hwang, J., Yang, R.: Common core standards the new us
  intended curriculum. Educational Researcher  40(3),  103--116 (2011)

\bibitem{rogers1983freedom}
Rogers, C.R.: Freedom to learn for the 80's. No. 371.39 R724f, Ohio, US:
  Merrill Publishing, 1983 (1983)

\bibitem{sinhaectel1}
Sinha, T., Bai, Z., Cassell, J.: A new theoretical framework for curiosity for
  learning in social contexts. In: 12th European Conference on Technology
  Enhanced Learning (2017)

\bibitem{sinha2015we}
Sinha, T., Cassell, J.: We click, we align, we learn: Impact of influence and
  convergence processes on student learning and rapport building. In:
  Proceedings of the 1st Workshop on Modeling INTERPERsonal SynchrONy And
  infLuence. pp. 13--20. ACM (2015)

\bibitem{sinha2015exploring}
Sinha, T., Zhao, R., Cassell, J.: Exploring socio-cognitive effects of
  conversational strategy congruence in peer tutoring. In: Proceedings of the
  1st Workshop on Modeling INTERPERsonal SynchrONy And infLuence. pp. 5--12.
  ACM (2015)

\bibitem{spektor2013science}
Spektor-Levy, O., Baruch, Y.K., Mevarech, Z.: Science and scientific curiosity
  in pre-school—the teacher's point of view. International Journal of Science
  Education  35(13),  2226--2253 (2013)

\bibitem{wu2013modeling}
Wu, Q., Miao, C.: Modeling curiosity-related emotions for virtual peer
  learners. IEEE Computational Intelligence Magazine  8(2),  50--62 (2013)

\bibitem{yin2012uspan}
Yin, J., Zheng, Z., Cao, L.: Uspan: an efficient algorithm for mining high
  utility sequential patterns. In: Proceedings of the 18th ACM SIGKDD
  international conference on Knowledge discovery and data mining. pp.
  660--668. ACM (2012)

\bibitem{zhao2016socially}
Zhao, R., Sinha, T., Black, A.W., Cassell, J.: Socially-aware virtual agents:
  Automatically assessing dyadic rapport from temporal patterns of behavior.
  In: International Conference on Intelligent Virtual Agents. pp. 218--233.
  Springer (2016)

\end{thebibliography}

\end{document}